# Quantitive evaluation of composition and biomolecular mapping of macrofungi spores by Raman spectroscopy

**Authors**: Petr Shvets[a*], Aleksandr Goikhman[b]

**Affiliation**: [a]Research and Educational Center "Functional Nanomaterials", Immanuel Kant Baltic Federal University, Aleksandra Nevskogo 14, 236041, Kaliningrad, Russian Federation

[b]Koenigssystems UG, 22869, Schenefeld bei Hamburg, Germany.

*corresponding author, email: pshvets@kantiana.ru

**Abstract.** Raman spectroscopy is widely applied to detect different chemical compounds in organic matter and create label-free high-resolution maps on the level of separate cells. The main advantage of this technique is the possibility to study objects in vivo without special pretreatment. However, it is rarely used to determine ratios between different substances or to perform deep quantitative analysis of the obtained spectra. In our study, we derive ratiometric equations that allow estimating the mass concentrations of triacylglycerols, proteins, sugars, polysaccharides, and DNA. We show that it is possible to determine average unsaturation and chain length from the spectra of lipids and concentrations of phenylalanine, tyrosine, and tryptophan from the spectra of proteins. We apply the derived equations to the Raman spectra of fungal spores of over 70 different species of macrofungi. We provide detailed characterization of lipids, proteins, and polysaccharides, which are contained in the spores. We believe that our study not only provides new fundamental knowledge in the field of mycology but also can become a basis for spectral quantification of any organic material, and our approach can be applied, for instance, to food diagnostics.

**Keywords:** Raman spectroscopy, Fungal spores, Triacylglycerols, Proteins, Saccharides, Raman mapping.

1. Introduction

Fungi are fascinating. For humans, they are beneficial in terms of food production [1] and medicine [2]; in nature, they are indispensable for biodegradation [3]. Spores are essential for the life cycle and reproduction of fungi, but they can also be prospective for certain medical, food, environmental, or energy applications [4]. For these purposes, it is important to know the content of bioactive components in the spore. It can be determined by different chemical methods, but they are usually complex and destructive, so vibrational spectroscopic techniques, which can be applied *in vivo*, become increasingly popular [5]. Among these, Raman spectroscopy is a very powerful tool enabling label-free, sub-micrometer mapping of the major compounds of living cells, including DNA, RNA, proteins, and lipids [6]. However, it is the least explored technique in the field of fungal research, and the most studies are limited to infrared spectroscopy characterization [5].

The spores of several wild-growing mushrooms were studied by Raman spectroscopy in 2001, but no assignment of the observed bands was provided [7]. The most detailed spectroscopic investigation of fungal spores was done by K. De Gussem et al. [8,9]. In the spores of the *Lactarius* genus, a high amount of lipids was detected; some Raman bands were assigned to chitin and amylopectin, and the possible presence of trehalose was mentioned [8]. In the spectra of *Collybia* and *Mycena* spores, some extra Raman peaks were related to nucleic acids and proteins [9]. Using linear discriminant analysis, it was possible to assign the spectra of spores to the correct genus but not on the species level [9]. Later, a similar technique was used to identify the spores of microfungi relevant to indoor contamination [10] and spores important for respiratory diseases [11]. Fungal filaments and fruiting bodies were studied by Fourier transform Raman spectroscopy, revealing the presence of chitin, *N*-acetylglucosamine, and (*R*)-glucan [12], or by surface-enhanced Raman spectroscopy resulting in the observation of nucleic acids or fungal pigments signals [13]. Fungal pigments were also investigated by several other groups [14,15].

The main disadvantage of the published works on Raman spectroscopy of fungi is the absence of simple ratiometric equations allowing calculating ratios between different components in the studied object (for instance, lipid to saccharide ratio in the spore of *Lactarius* fungus) or some parameters of a certain detected substance (for instance, lipid unsaturation). Also, in most cases, the studied spectra are limited to the fingerprint region ($k < 1800$ cm$^{-1}$), so valuable information about C–H bonds might be omitted. It could be one of the reasons for a not full / not fully correct interpretation of the observed spectra. Finally, the world of fungi is very large and diverse (~0.15 described and more than 10 million estimated species [16]), so only a tiny fraction of this world has been currently characterized by Raman spectroscopy.

In our work, we develop simple approaches to obtain quantitative information about organic substances contained in fungal spores based on their Raman spectra accurately measured in the whole spectral range (100 – 3500 cm$^{-1}$). We use these approaches to describe the composition of the spores of over 70 different species and to build compositional maps of some individual large spores. Our equations can be applied not only to fungal-related samples but also to any organic material requiring diagnostics.

2. **Material and methods**

We study the spores of wild-growing macrofungi. The current study is limited to white or lightly-colored spores, which do not produce strong fluorescence during Raman measurement. All samples were collected in the Kaliningrad region, Russian Federation. The details about the collecting date and place for each sample can be found in Supporting Information, Appendix I. The fresh mushroom was placed on monocrystalline silicon coated by a vanadium film. Such a substrate is Raman-inactive and thus it is beneficial for studying small semi-transparent objects (it was used for the same purpose in our previous work [17]). After several hours, the spores are deposited on the substrate and are investigated by Raman spectroscopy without further preparations. Prior to scanning electron microscopy (SEM), the spores were magnetron-coated by a thin platinum film to improve the resolution.

Raman scattering study was done using a Horiba Jobin Yvon micro-Raman spectrometer LabRam HR800 with an x100 magnification objective (numerical aperture of 0.9). Measurements were conducted at room temperature in the air environment. A He-Ne laser with a 632.8 nm wavelength was used to excite Raman scattering. The laser power on the sample in most cases was 5 mW, and the spot diameter was about 2 μm. Some spores were unstable under such conditions, so we had to decrease the power to exclude sample degradation.

SEM images were acquired in the JEOL JSM-6390LV microscope. We used an accelerating voltage of 30 kV and a secondary electron detector. The work distance was about 6 mm.

3. **Results**

3.1. Lipid-based spores.

Raman spectra of spores of some higher basidiomycetous fungi are mostly comprised of the Raman bands of lipids [8]. We will start our discussion with a typical spore of this type belonging to *Tricholomopsis decora*. The Raman spectrum of the spore (Figure 1a) is indeed nearly identical to the spectrum of olive oil (Figure 1b), confirming that the spore mainly consists of triacylglycerols (TAGs). In TAGs, Raman spectroscopy can be used to determine the average saturation and chain length [18,19]. To do this, it is necessary to measure the intensity of several peaks: ~1260 cm$^{-1}$ (corresponding to the =C−H deformation), ~1440 cm$^{-1}$ (CH$_2$ scissoring vibrations), ~1660 cm$^{-1}$ (C=C stretching vibrations), ~2850 cm$^{-1}$ (symmetric stretching of C–H in CH$_2$), ~2930 cm$^{-1}$ (chain-end CH$_3$ symmetric stretching), and ~3000 cm$^{-1}$ (=C−H stretching mode) [20]. However, published ratiometric relations or components used for principal component analysis should be used with caution because they can depend on the instrumentation [21]. Our Raman spectrometer is equipped with the 632.8 nm He-Ne laser, and for the monocrystalline silicon reference sample, it produces the spectral line at ~520 cm$^{-1}$ with full width at half maximum (FWHM) of ~3.8 cm$^{-1}$. It is in perfect agreement with one of the modes used for TAG identification in ref. [21]. Thus, we will use the relations from this

publication:

$$N_{C=C}^{(1)} = (2.3 \pm 0.3) I_{1262}/I_{1448} + (-0.2 \pm 0.4) \quad (1)$$

$$N_{CH_2} = (5.2 \pm 0.7) I_{2850}/I_{2935} + (5.9 \pm 1.5) \quad (2)$$

$$N_{C=C}^{(2)} = N_{CH_2}((0.047 \pm 0.002) I_{1655}/I_{1448} + (-0.02 \pm 0.01)) \quad (3)$$

$$N_{C=C}^{(3)} = 0.5 * N_{CH_2}((0.76 \pm 0.09) I_{3005}/I_{2850} + (0 \pm 0.05)) \quad (4)$$

where $I_{xxx}$ is the intensity of Raman signal at $xxx$ cm$^{-1}$ after baseline correction, $N_{C=C}$ is the number of double bonds, and $N_{CH2}$ is the number of CH$_2$ groups (the total chain length can be calculated as $N_C = 2 + N_{CH2} + 2 * N_{C=C}$).

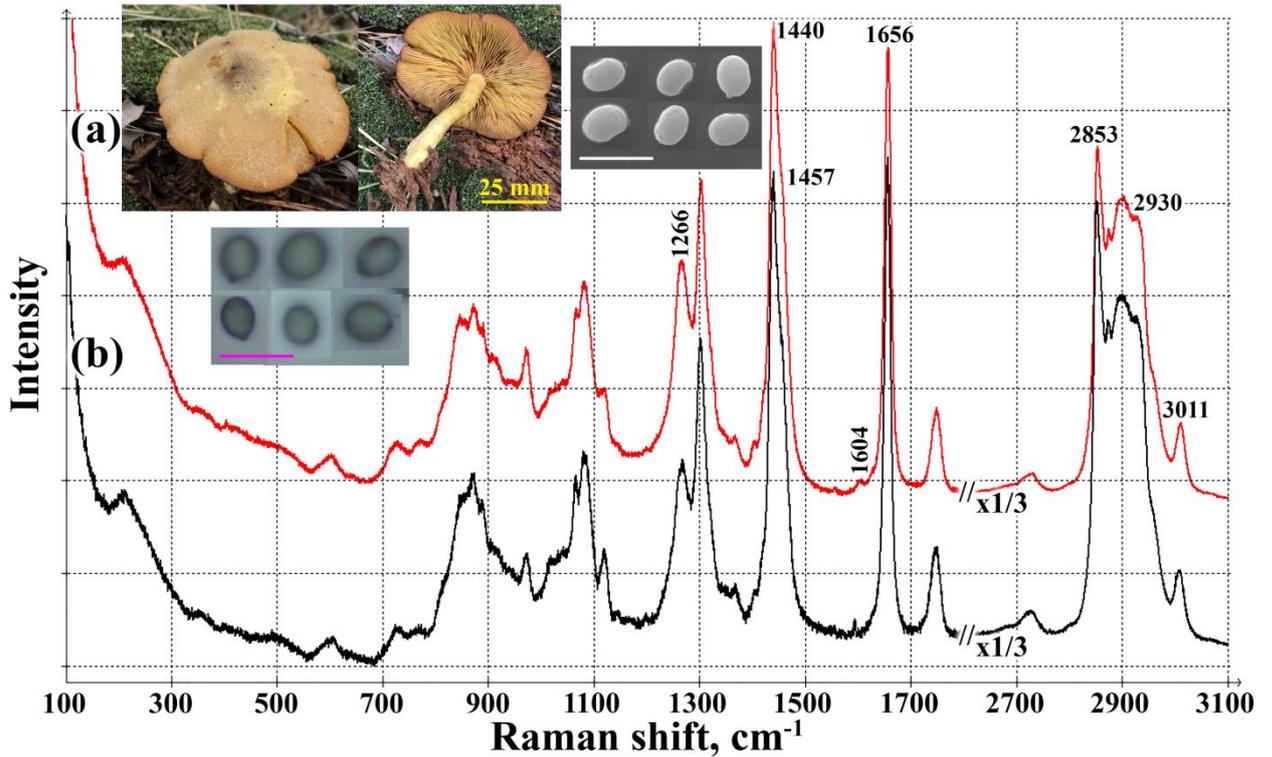

Figure 1. Raman spectra of (a) spore of *Tricholomopsis decora* and (b) olive oil. Raman intensity in the C–H region ($k > 2600$ cm$^{-1}$) was divided by 3. Numbers near the peaks are the corresponding wavenumbers (in cm$^{-1}$). Insets are the photograph of the fungus, optical image of its spores as seen through Raman microscope objective, and SEM image of the spores. Scale bars on the microimages are 10 μm.

For olive oil, using these equations, we can obtain $N_C = 17.0 \pm 1.8$, $N_{C=C} = 0.8 \pm 0.3$. For *Tricholomopsis decora*, the values are almost the same: $N_C = 16.7 \pm 1.8$, $N_{C=C} = 1.0 \pm 0.3$, confirming the similarity of the corresponding TAGs. Except for the Raman bands of TAG, the only weak extra peak present in the spore spectrum is located at 1604 cm$^{-1}$. This peak likely indicates the presence of pulvinic acid-type compounds, probably atromentic acid, one of the very common fungal pigments [14].

3.2. Protein-based spores.

During our investigations, we discovered that spores of some fungi produce Raman spectra, which have no visible traces of TAGs. Instead, they are mostly comprised of the Raman bands of proteins. A typical spore of this type belongs to *Gymnopus dryophilus* (Figure 2a). For comparison, we give the spectrum of the reference protein (from hen egg white, Figure 2b).

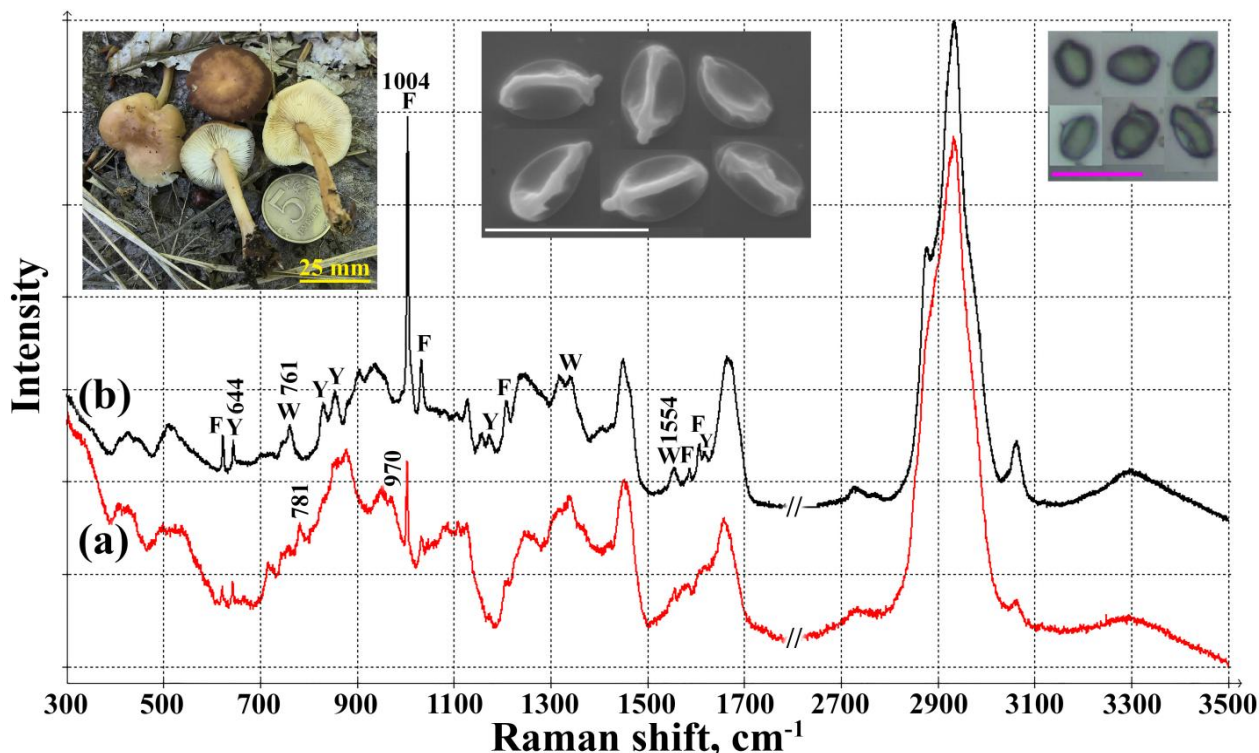

Figure 2. Raman spectra of (a) spore of *Gymnopus dryophilus* and (b) egg white protein. Letters F, Y, and W mark the modes of phenylalanine, tyrosine, and tryptophan, respectively. Insets are the photograph of the fungus, optical image of its spores as seen through Raman microscope objective, and SEM image of the spores. Scale bars on the microimages are 10 μm.

Overall, the spectra are quite similar, though there are some pronounced differences in the shapes of C–H bands (2800–3100 cm$^{-1}$) and low-wavenumber bands (<600 cm$^{-1}$). To address these differences, we will analyze the spectrum of the spore of *Peziza Varia* (Figure S1a (here and elsewhere, Figure S*xx* refers to Supporting Information)). This spectrum has well-defined peaks in the low-wavenumber region, and their positions and relative intensities are in good agreement with those for trehalose dissolved in water (Figure S1b). Note that Raman spectra of saccharides strongly depend on their crystalline state [22], so our results suggest that sugars in fungal spores are amorphous. Now, we can adjust the intensities of Raman spectra of *Peziza Varia* spore and trehalose (in the range 390 – 560 cm$^{-1}$; it is done in Figure S1 and Figure S2a,b) and calculate the difference between two spectra in the region of C–H vibrations (Figure S2c).

We can see that this difference is very similar to the spectrum of egg white (Figure S2d). A shoulder at ~2850 cm$^{-1}$ is caused by the presence of lipids (Figure S2e).

Now, we can estimate protein to sugar content in the spores. To do this, we created two similar solutions (12 wt.% trehalose in water and 12 wt.% protein in water (natural composition of hen egg white)) and measured them under the same acquisition conditions. The results are given in Figure S3a,b. The spectrum of the *Peziza Varia* spore, $S(k)$ (Figure S3c), is considered as a linear combination of the obtained spectra of trehalose, $S_t(k)$, egg protein, $S_e(k)$, and a small contribution of lipid (*Tricholomopsis decora*, $S_l(k)$, Figure S3d):

$$S(k) = K * S_t(k) + L * S_e(k) + M * S_l(k) \tag{5}$$

where $K$, $L$, and $M$ are constants. The best fit is achieved for $K = 3.85$, $L = 15.26$, and $M = 0.061$, so protein to trehalose ratio is P : T = 0.8 : 0.2. Note that the shape of sugar-protein contribution to the Raman spectrum in C–H region is the same for all protein-containing spores (e.g. compare Figures S2a and S2g), so this ratio is likely universal.

Distinct Raman lines of trehalose are rarely seen in the spectra of spores. Apart from *Peziza Varia*, we detected trehalose in *Phyllotopsis nidulans*. We studied two different samples: a fresh one (Figure S4) and one that survived at least one month in the winter forest. The "winter" sample (Figure S5a) has broad weak Raman peaks at ~408 and 435 cm$^{-1}$, indicating that sugars might still be related to trehalose. However, all other characteristic trehalose peaks are missing. The same is observed in *Gymnopus dryophilus* (Figure S5d) and many other spores with high protein content. We believe that trehalose is the main sugar component in all these cases, but its Raman response is strongly affected by the interaction with protein.

In proteins, Raman spectroscopy can be used to obtain information about secondary structures and amino acid residues [23]. Among different amino acids, only those with aromatic side chains can be easily distinguished: phenylalanine (the strongest narrow peak at 1003 cm$^{-1}$ (ring breathing mode) and weaker modes at 622 (phenyl ring breathing vibrations), 1032 (CN stretching), 1207 (side chain vibrations), 1586 (phenyl ring bond-stretching vibrations (out-of-phase)), and 1606 cm$^{-1}$ (phenyl ring bond-stretching vibrations (in-phase)), tyrosine (643 (C–C ring twist), 830 (Y6 mode), 850 (Y5 mode), 1177 (Y4 mode), and 1613–1616 cm$^{-1}$ (Y1 mode)), and tryptophan (757 (coupled vibrations of in-phase breathings of benzene and pyrrole), 1340 (W7 mode), and ~1550 cm$^{-1}$ (W3 mode)). For quantitative estimations, we will use simple ratiometric equations:

$$C_F = 4.64\% * S * (I^s_{1004}/I^s_{2934})/(I^e_{1004}/I^e_{2934}) = 8.23\% * I^s_{1004}/I^s_{2934} \tag{6}$$

$$C_Y = 3.08\% * S * (I^s_{644}/I^s_{2934})/(I^e_{644}/I^e_{2934}) = 58.5\% * I^s_{644}/I^s_{2934} \tag{7}$$

$$C_W = 1.01\% * S * (I^s_{761}/I^s_{2934})/(I^e_{761}/I^e_{2934}) = 11.8\% * I^s_{761}/I^s_{2934} \tag{8}$$

where *S* is a coefficient taking into account sugar contribution to the intensity at 2934 cm$^{-1}$ ($S$ = 1.3), $C_F$, $C_Y$, and $C_W$ are mass concentrations of phenylalanine, tyrosine, and tryptophan in crude protein, respectively, $I^s_{xxx}$ and $I^e_{xxx}$ are intensities of Raman signal at *xxx* cm$^{-1}$ after baseline correction for the spore (after the removal of lipid part) and reference egg white protein, respectively, and concentrations of amino acids in egg (4.64 ± 0.17 %, 3.08 ± 0.09 % and 1.01 ± 0.08 %) were taken from ref. [24]. For *Gymnopus dryophilus*, using these equations, we can obtain $C_F$ = 2.54 ± 0.15 %, $C_Y$ = 4.5 ± 1.1 %, and $C_W$ = 0.63 ± 0.17 % (uncertainties were estimated using signal-to-noise ratio).

The secondary conformation of proteins can be analyzed using the amide I band (1600–1700 cm$^{-1}$, C=O stretching vibrations with small contribution from out-of-plane C–N stretching). However, such analysis requires many-component deconvolution, which can not be unambiguously done for a relatively weak band with potential contributions from unidentified components. Thus, we will not try to derive any quantitative estimations for the secondary structure of proteins. The example for amide I band deconvolution can be found in Figure S6. For the fitting, we tried to use a minimal number of components. The components at 1652/1657 cm$^{-1}$ and 1673/1677 cm$^{-1}$ can be assigned to α-helix and β-turns (or random structures) conformations, respectively [23]. From such a deconvolution, we can conclude that proteins in the spore of *Gymnopus dryophilus* have a significantly higher amount of alpha-helical structure than proteins in hen egg white.

Raman spectroscopy is an efficient tool for probing sulfide bonds. S–H stretching in cysteine is observed at 2565–2585 cm$^{-1}$, and the S–S stretching band is located at ~ 508, 525, or 544 cm$^{-1}$, depending on the conformation of the disulfide bond [23]. However, there are no well-defined peaks at these wavenumbers, so the concentration of cysteine in our samples is below the detection limit.

Apart from saccharides and protein peaks, the Raman spectrum of *Gymnopus dryophilus* contains an extra feature at 781 cm$^{-1}$. In cells, the peak at ~780 cm$^{-1}$ is considered as a marker of DNA [25], because it is associated with the breathing mode of polyimide ring present in cytosine and uracil [26,27]. To quantify the DNA / protein ratio ($C_{DNA}$ / $C_P$), we used reference spectra of DNA and CT-histone (Figure S7a and Figure S7b, respectively, [28]). First, we normalized the spectrum of *Gymnopus dryophilus* to the published spectrum of CT-histone (using peaks at ~1450 and 1650 cm$^{-1}$, Figure S7c) and then derived the following relation:

$$C_{DNA}/C_P = 1 * F * (I_{781}/I_{2934})/(I^R_{787}/I^R_{2934}) = 1.45 * I_{781}/I_{2934} \qquad (9)$$

where $I_{xxx}$ is the intensity of Raman signal at *xxx* cm$^{-1}$ after baseline correction, $I^R_{787}$ and $I^R_{2934}$ are measured for the reference DNA and for the normalized spectrum of the *Gymnopus* *dryophilus* (after subtracting lipid component), and coefficient *F* takes into account the different resolution of Raman spectrometers by comparing FWHM for DNA lines (*F* ~ 0.5). The calculation for the spore gives $C_{DNA} / C_P = 0.080 \pm 0.015$ (uncertainty was estimated using signal-to-noise ratio).

### 3.3. Lipid-protein mixed spores.

In most cases, Raman spectra of spores contain bands of both lipids and proteins. A typical spore of this type belongs to *Limacella illinita* (Figure 3a). To analyze such spectra, we first had to obtain normalized reference spectra of lipid and protein. Normalized protein, $S_p(k)$ (in fact, we will further use a mixed sugar-protein spectrum, $S_{sp}(k)$), can be derived from Equation (5) as:

$$S_{sp}(k) = \frac{0.2 S_t(k) + 0.8 S_p(k)}{1} \approx \frac{K * S_t(k) + L * S_e(k)}{K + L} = \frac{S(k) - M * S_l(k)}{K + L} \quad (10)$$

where $S(k)$ and $S_l(k)$ are the spectra of *Peziza Varia* and *Tricholomopsis decora* spores, $S_t(k)$ and $S_e(k)$ are the spectra of reference 12wt% solutions of trehalose and egg protein, $K = 3.85$, $L = 15.26$ and $M = 0.061$. To obtain a normalized lipid spectrum, we created another reference solution (12 wt.% sunflower oil in acetone). We measured the Raman response of this solution under the same conditions as before and extracted the signal from the lipid. Finally, we normalized the spectrum of *Tricholomopsis decora* to the obtained signal. Normalized spectra are given in Figure 3b,c. Our data were compared with those obtained for lipoproteins extracted from the blood plasma [29], and there is a good agreement between normalized intensities despite the differences in the nature of the studied objects (Figure S8). The spectrum of the lipid-protein mixed spore, $S(k)$, is considered as a linear combination of the normalized spectra of lipid (*Tricholomopsis decora*, $S_l(k)$) and protein, $S_{sp}(k)$:

$$S(k) = A * S_l(k) + B * S_{sp}(k) \quad (11)$$

where *A* and *B* are constants, and only the C–H Raman region is analyzed (2800–3100 cm$^{-1}$). For *Limacella illinita*, $A_{LI} = 1.4$, $B_{LI} = 2.4$, so protein to lipid ratio P : L = 0.63 : 0.37 (note that the crude protein content is 0.63*0.8 = 0.5, while the remaining part, 0.13, is sugars). The corresponding linear combination is given as a black line in Figure 3, and we can see a perfect agreement between the two spectra (for $k < 3000$ cm$^{-1}$). The peak at 3012 cm$^{-1}$ is not well-fitted because of the different saturation of lipids in *Limacella illinita* and *Tricholomopsis decora*. To estimate lipid saturation, we used the full spectrum of *Gymnopus dryophilus* (in the fingerprint region, *Peziza Varia* contains unusually strong characteristic trehalose bands and some extra unidentified Raman peaks, so it is not good for batch analysis). First, we used equation (11) to obtain $A_{GD} = 0.24$ and $B_{GD} = 7.17$ (lipid content is ~3%). Normalized protein spectrum (in the whole range of *k*) is calculated as:

$$S_{sp}(k) = (S_{GD}(k) - A_{GD} * S_l(k))/B_{GD}$$

Now, we can multiply this by $B_{LI}$ and subtract it from the spectrum of *Limacella illinita* to obtain the lipid part (the result is given in Figure 3d). After that, we used equations (1–4), resulting in $N_C = 17.6 \pm 1.7$, $N_{C=C} = 1.5 \pm 0.3$. Note that the lipid part of *Limacella illinita* contains no visible peaks of amino-acid signals. It means that the amino-acid profiles of *Limacella illinita* and *Gymnopus dryophilus* are similar. Indeed, direct calculation based on the spectrum gives $C_F = 2.70 \pm 0.19$ %, $C_Y = 4.9 \pm 1.1$ %, and $C_W = 1.1 \pm 0.2$ %.

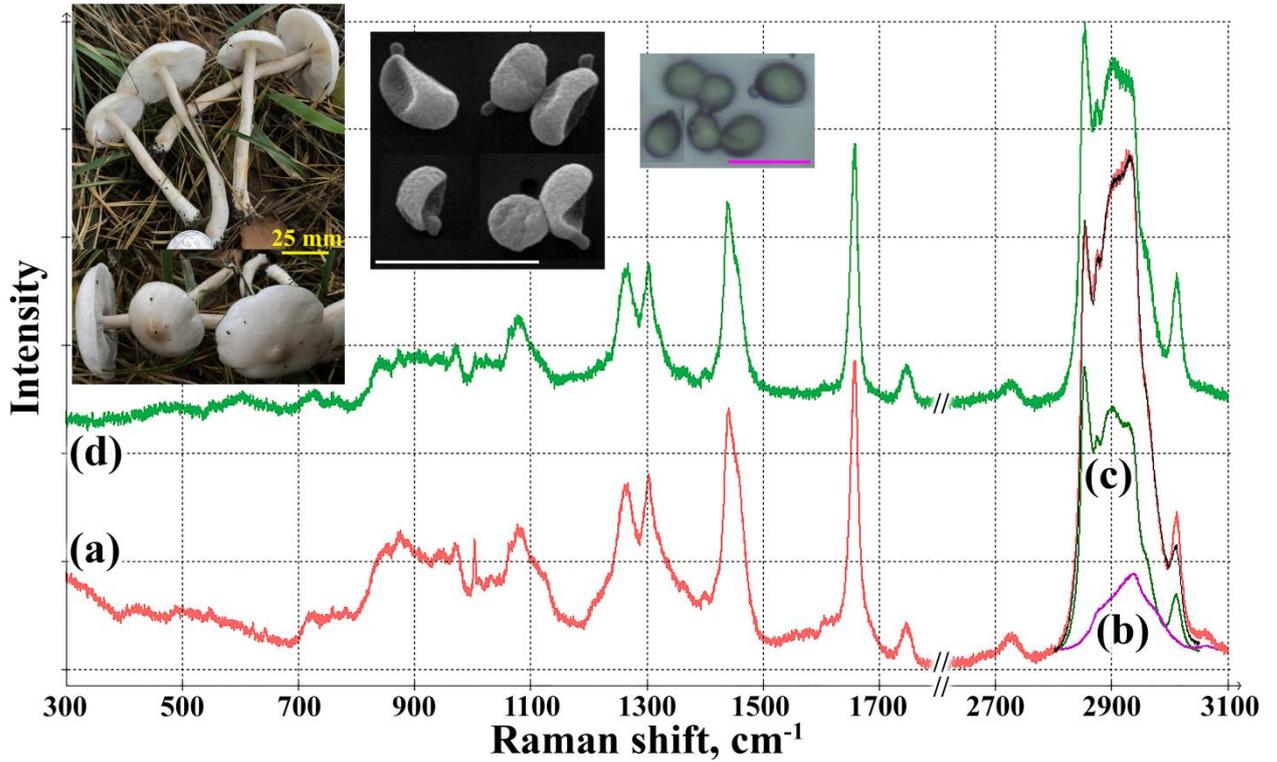

Figure 3. Raman spectra of spore of (a) *Limacella illinita*, (b) *Peziza Varia* without lipid contribution, and (c) *Tricholomopsis decora*. The thin black line is a linear combination of spectra (b) and (c). (d) Raman spectrum of spore of *Limacella illinita* after excluding protein component. Insets are photograph of the fungus, optical image of its spores as seen through Raman microscope objective, and SEM image of the spores. Scale bars on microimages are 10 µm.

3.4. Hydrocarbon-containing spores.

Amyloid spores of *Russula* and *Lactarius* genera contain Raman-detectable amounts of glucans [9]. A readily available source of glucans (amylopectin) is corn starch. Unfortunately, starch is insoluble in simple solvents, so we had to develop a special procedure to obtain its normalized spectrum. First, we put 3 wt.% of starch in water. Under constant stirring, we started adding NaOH into the colloidal solution. When NaOH concentration reached about 1 M, the solution turned clear, indicating the successful dissolution of starch in accordance with the literature [30]. We waited several hours for all the bubbles to go out and then collected the spectrum of the solution (Figure S9a). We normalized the obtained spectrum, taking into account different solution concentrations and acquisition times (compared to the values used for obtaining Figure 3 b,c). Comparing the spectra of the solution and powder starch (Figure S9b), we can see that the spectra are generally the same, but amorphization shifted the first characteristic Raman peak to 489 cm$^{-1}$ and made it broader. In fungi, the position of the glucan peak is ~480 cm$^{-1}$, indicating that it is semi-crystalline. Thus, for the spore quantification, we normalized the spectrum of powder starch in a way that the integral intensities of the peaks at 480 cm$^{-1}$ (Figure S9b) and 489 cm$^{-1}$ (Figure S9a) were identical. Finally, we can calculate:

$$D = I_{480}/I^R_{480} \qquad (12)$$

where $I_{480}$ and $I^R_{480}$ are the intensities of Raman signals at 480 cm$^{-1}$ after baseline correction for the spore and the normalized reference spectrum of corn starch powder, respectively. Now, we can subtract the starch spectrum multiplied by $D$ from the spectrum of spore and find coefficients $A$ and $B$ from equation (11). The ratios between $A$, $B$, and $D$ are proportional to the mass concentrations of the corresponding substances. For example, for *Russula sanguinea* (Figure S9c), we can obtain lipid : protein/sugar : glucan = 0.48 : 0.4 : 0.12.

We should note that the differences between Raman spectra of glucans (amylopectin, amylose, cellulose) are minimal [8,31], so based on only two Raman lines usually visible in spores (~479 and ~943 cm$^{-1}$), it is not possible to claim that a certain glucan is present in the fungus. Incubation with β-amylase was used to determine that fungal starch consists only of short-chain amylose molecules (*Russula* sp. and *Clavicorona pyxidata* (*Artomyces pyxidatus*) were among the studied samples) [32]. Thus, the observed Raman spectra should be interpreted as containing amylose, not amylopectin, as was stated before [9]. In *Artomyces pyxidatus*, we indeed detected a significant amount of glucans (~8%, Figure S10). In the Raman spectrum, we can see an extra peak at 1127 cm$^{-1}$, indicating that the spore contains amylose (the corresponding peak of amylopectin is at 1131 cm$^{-1}$) [31]. A much higher concentration of glucans can be found in *Auriscalpium vulgare* (Figure S11, ~22 wt. %). The spore is protein-based (lipid concentration is just ~4 wt. %), so it is easier to observe lower-intensity glucan peaks. The positions of some of them (578, 1083, and 1128 cm$^{-1}$) further confirm that the spectrum comes from amylose [31].

We studied several fungi from the *Lactarius* genus and confirmed that their spores contained amylose (A, 12 – 15 wt. %). The studied samples significantly varied in terms of protein to lipid ratio (P : L): *L. pubescens* (Figure S12, P : L : A = 0.29 : 0.59 : 0.12), *L. deliciosus* (Figure S13, P : L : A = 0.1 : 0.76 : 0.14), *L. quietus* (Figure S14a, P : L : A = 0.32 :

0.53 : 0.15), and *L. aurantiacus* (Figure S14b, P : L : A = 0.42 : 0.44 : 0.14). Some spores also contained pulvinic acid-type compounds, probably with different structures (peaks at 1604 and 1606 cm$^{-1}$ for *L. quietus* and *L. deliciosus*, respectively).

In the *Russula* genus, spores contain 4 – 13 wt. % of amylose and a variable amount of proteins. We measured the spectra of *R. claroflava* (Figure S15, P : L : A = 0.08 : 0.79 : 0.13), *R. cyanoxantha* (Figure S16, P : L : A = 0.12 : 0.82 : 0.06 and another sample in Figure S17a, P : L : A = 0.08 : 0.83 : 0.09), *R. nigricans* (Figure S17b, P : L : A = 0.16 : 0.80 : 0.04), *R. aeruginea* (Figure S17c, P : L : A = 0.27 : 0.62 : 0.11), and *R. vesca* (Figure S17d, P : L : A = 0.27 : 0.64 : 0.09). We have also detected glucans in *Melanoleuca melaleuca* (Figure S18, ~11 wt. %).

## 4. Discussion.

Equations (1–12) were used to quantify 71 fungi species with light-colored spores. The complete set of data can be found in Supporting Information (Table S1 and Figures S9-S48). In most cases, the obtained spectra can be treated as a superposition of two reference spectra for lipid and protein, just like we did with *Limacella illinita*. Equation (11) works very well for almost all studied species regardless of the lipid unsaturation (see Figure S49 for some examples). It means that the amount of CH$_2$ groups does not depend on the amount of double bonds because equation (2) is, in fact, always applied to the same reference spectrum. Thus, longer lipids in fungal spores have more double bonds. From our data, in most cases, $N_C$ ~ 17 and $N_{C=C}$ ~ 1. In a very special case of *Laetiporus sulphureus* (Figure S41b), $N_C$ ~ 15 and $N_{C=C}$ ~ 0. In fungi, we can expect the presence of C16:0 palmitic acid, C18:1 cis-oleic acid, and C18:2 cis-linoleic acid [33,34]. So, it seems that equation (2) underestimates the number of CH$_2$ groups (by 1). It also explains the discrepancy we observed for the reference olive oil (we obtained $N_C$ ~ 17 instead of 18). Taking this into account, we can conclude that lipids in *Laetiporus sulphureus* spores mainly consist of palmitic acid, while spores of the majority of species contain oleic acid (e.g. *Tricholoma terreum* with $N_{C=C}$ ~ 1.0 and $N_C$ ~ 16.8 ($N_C$ + 1 ~ 18)). In rare cases, lipids contain more than one double bond (e.g. *Limacella illinita*, $N_{C=C}$ ~ 1.5 and $N_C$ ~ 17.6, *Echinoderma asperum* $N_{C=C}$ ~ 1.7 and $N_C$ ~ 17.9, or *Sarcoscypha austriaca* $N_{C=C}$ ~ 1.8 and $N_C$ ~ 18.9). Such lipids should have longer chains and probably contain significant amounts of rare C20:2 eicosadienoic acid. On the other hand, formal uncertainty in the determination of $N_C$ by Raman spectroscopy is large (~ ±2), so it is possible that these spores just contain larger concentrations of linoleic acid.

Sometimes, equation (11) gives a rather bad fit. An example is *Laccaria laccata* (Figure S50), indicating that its spores contain some special lipids. It is confirmed by the presence of two shoulders (1633 and 1674 cm$^{-1}$, Figure S39a) on ~1660 cm$^{-1}$ peak corresponding to C=C stretching vibrations. A very special case is the spectrum of *Cantharellus cibarius*, which contains a peak at 2216 cm$^{-1}$ (and a smaller one at 2253 cm$^{-1}$, Figure S48a). Such wavenumbers (and a doublet structure) are typical for triple C≡C bond stretching vibrations in acetylenic fatty acids [35]. The position of the stronger band (2216 cm$^{-1}$) suggests that a triple bond should be located in the vicinity of other unsaturated groups [35]. This is in agreement with the structure of acetylenic acids isolated from the *Cantharellus cibarius* [36]. Moreover, the position of another C=C stretching peak at 1619 cm$^{-1}$ also supports the occurrence of conjugated arrangements of triple and double bonds [37]. The spore of *Pseudocraterellius undulatus* contains the same acetylenic acids (Figure S48b), but their concentration is much higher revealing many weaker Raman features of these unusual substances.

There are three equations ((1), (3) and (4)) for estimation of $N_{C=C}$. Two of them give nearly identical, reasonable results: $N^{(1)}_{C=C} \approx N^{(3)}_{C=C}$, while equation (3) gives significantly underestimated values (see Figure S50 for correlation plots). That is why $N_{C=C}$ column in Table S1 contains the average value $(N^{(1)}_{C=C} + N^{(3)}_{C=C})/2$, disregarding the results for $N^{(2)}_{C=C}$. However, equation (3) can still be used if it is modified by a factor of 2.3:

$$N^{(2\prime)}_{C=C} = 2.3 * N_{CH_2}((0.047 \pm 0.002) I_{1655}/I_{1448} + (-0.02 \pm 0.01)) \tag{3'}$$

The protein-to-lipid ratio in the spore varies significantly from species to species, including fungi belonging to the same genus. For instance, lipid concentration in *Tricholoma scalpturatum* is ~38%, in *Tricholoma stiparophyllum* – ~58%, in *Tricholoma terreum* – ~71%, and in closely related *Tricholomopsis decora* – ~100%. Thus, if identification of a spore (on species level) is required based on a Raman spectrum, it is necessary to measure and analyze (using Equation (11)) the region of C–H bonds (2800–3100 cm$^{-1}$). However, this approach would require preliminary careful measurements of several spores of several different samples to estimate variations and average values for each of the studied species.

Sometimes, Raman spectra of the spores contain extra notable peaks. Unfortunately, it is not always easy to identify them exactly. *Clitocybe nebularis* (Figure S24) has a strong peak at 856 cm$^{-1}$. The strongest band in this region might belong to α-glycosidic bonds in saccharides [38]. The same saccharide peak was detected in *Hymenopellis radicata* (Figure S31b) and *Lentinus tigrinus* (Figure S32). Some spores contain pulvinic acid-type pigments (peak at ~1604 cm$^{-1}$): *Boletus reticulatus* (Figure S19a), *Tylopilus felleus* (Figure S19b), *Boletus edulis* (Figure S19c), *Hortiboletus bubalinus* (Figure S20), *Clitocybe nebularis* (Figure S24), *Gymnopus peronatus* (Figure S30a), *Mycena galericulata* (Figure S31a), *Hymenopellis radicata* (Figure S31b), *Pluteus cervinus* (Figure S37), and *Polyporus tuberaster* (Figure S41a). *Lepiota clypeolaria* (Figure S29b) and *Chlorophyllum rhacodes* (Figure S38a) have a peak at 895 cm$^{-1}$. It might indicate that the protein in the spores contains a significant amount of glycine [39]. *Neolentinus lepideus* (Figure S33) has a strong peak at 1550 cm$^{-1}$, which is likely due to the presence of a fungal pigment grevillin B [14]. The same pigment can explain the doublet around 500 cm$^{-1}$, but there are also some extra unattributed peaks in the fingerprint region. *Pleurotus* sp. (Figure S36) have peaks at 545-547 and 1452 cm$^{-1}$, but their origin is uncertain. The peak at 1045 cm$^{-1}$ in *Panellus stipticus* (Figure S40) might be associated with ring-breathing mode in pyridine derivatives [40].

Some spores are much bigger than the focused laser beam, so we can measure Raman spectra from different points in the spore and investigate the distribution of organic components. In most cases, the variations are minimal (the example is *Neolentinus lepideus*, Figure S33). In some cases, however, it is possible to obtain informative maps showing the distribution of main components (lipids and proteins). A good example is *Sarcoscypha austriaca*, Figure 4, which demonstrates the variation of lipid to protein ratio from 6 to 34%. Less evident variation can be seen in the map of *Tremella mesenterica* spore (from 23 to 42%, Figure S52).

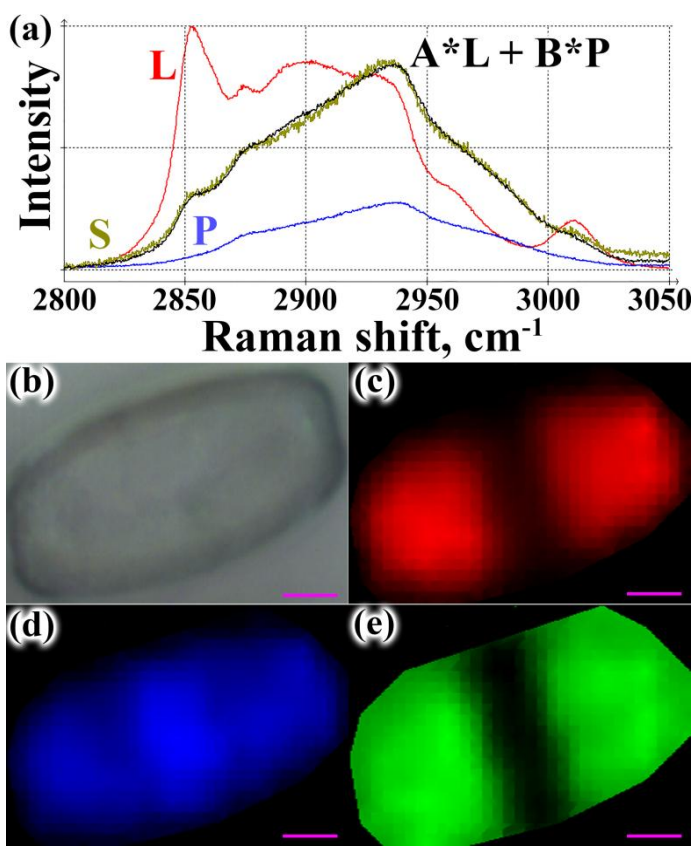

Figure 4. (a) Raman spectrum collected in the center of spore of *Sarcoscypha austriaca* (S) with normalized spectra of lipid (L) and protein (P). The spectrum S is fitted by the linear combination of two reference spectra: S = A*L + B*P (A = 0.17, B = 2.64). (b) Optical image of the spore. (c-e) Raman maps, showing the distribution of lipid (c), protein (d), and lipid-to-protein ratio (e) (color intensities are proportional to A, B, and A/(A+B), respectively). Scale

bars are 5 μm.

## 4. Conclusions

Raman spectroscopy can be used to obtain rich data about the chemical composition of organic objects. We investigated spores of over 70 different species of macrofungi growing in the region of the Baltic Sea. The main components of spores are lipids (triacylglycerols) and proteins. We developed a simple approach to the determine relative mass concentration of these components and showed that it can vary within a wide range depending on the fungi species (from 1 : 0 to almost 0 : 1). Large individual spores can be mapped, showing non-uniform distribution of lipid to protein ratio.

Using simple ratiometric equations, it is possible to estimate the average chain length and unsaturation of triacylglycerols. We showed that in most cases, spores contain C18:1 cis-oleic acid. Sometimes, the spectrum is dominated by C16:0 palmitic acid (*Laetiporus sulphureus*) or contains triacylglycerols with a higher degree of unsaturation (*Limacella illinita*, *Echinoderma asperum*, or *Sarcoscypha austriaca*). In *Cantharellus cibarius* and *Pseudocraterellius undulatus*, Raman spectroscopy detects the presence of acetylenic fatty acids.

Proteins in the spores are associated with sugars, and in some cases, we could detect the presence of amorphous trehalose (*Peziza Varia* and *Phyllotopsis nidulans*). It seems that, normally, trehalose interacts with protein, losing its characteristic Raman bands in the fingerprint region but keeping its contribution to the spectrum in the C–H region. The sugar content in the spore seems to be universal, ~20% of the mass of the protein. Simple ratiometric equations allow calculating the mass concentrations of phenylalanine, tyrosine, and tryptophan in crude protein. Protein is also linked with DNA. DNA concentration can be estimated by the intensity of the Raman peak at 781 cm$^{-1}$, yielding 5 – 10% with respect to the mass of protein.

Some spores contain polysaccharides (amylose) visible as Raman peak at ~480 cm$^{-1}$. We could detect 5–15 wt% of amylose in *Artomyces pyxidatus*, *Melanoleuca melaleuca*, *Xeromphalina campanella*, *Russula*, and *Lactarius* sp. The highest concentration of amylose (25%) was found in protein-based spores of *Auriscalpium vulgare*.

## 5. Acknowledgements

This work was prepared with support from the Ministry of Science and Higher Education of the Russian Federation (project FZWM-2024-0011).

**Data availability**

The data that support the findings of this study can be found in Supporting Information.